\documentclass{PoS}

\title{Heavy flavor spectroscopy at the Tevatron}

\ShortTitle{HQL 2010}

\author{\speaker{Diego Tonelli}\thanks{for the CDF and D\O\ Collaborations.}\\
        Fermilab\\ P.O. Box 500, Batavia, IL, 60510 -- USA \\
        E-mail: \email{tonel@fnal.gov}}

\abstract{During the last decade, the CDF and D\O\ experiments at the Tevatron have been pursuing an extensive heavy flavor spectroscopy program that provides significant advancements in the understanding of masses of ground state and orbitally excited $b$--hadrons,  and phenomenology of exotic $XYZ$ states. It also yields a breakthrough in our knowledge of bottom baryons with the first observation of the $\Sigma_b^{\pm(*)}$, $\Xi_b^-$,  and $\Omega_b^-$ families. After briefly reviewing previous results, I will focus on the latest CDF measurements of $\Sigma_b^{\pm(*)}$ baryons and exotic $J/\psi\phi$ states, both updated to 6 fb$^{-1}$ of data, and on a new analysis of charmed baryons.}

\FullConference{The Xth Nicola Cabibbo International Conference on Heavy Quarks and Leptons,\\
		October 11-15, 2010\\
		Frascati (Rome) Italy}

\def \rightdownarrow
 {\kern.3em
 \rule[.5ex]{.15mm}{2ex}
 {\mbox{$\kern-0.1em{\longrightarrow}$}}      % down-right arrow for cascade decay equations
 }

\def\lessim{\mathrel {\vcenter {\baselineskip 0pt \kern 0pt  % minore/circa uguale
\hbox{$<$} \kern 0pt \hbox{$\sim$} }}}

\def\gessim{\mathrel {\vcenter {\baselineskip 0pt \kern 0pt   % maggiore/circa uguale
\hbox{$>$} \kern 0pt \hbox{$\sim$} }}}

\newcommand{\lumifb}{\mbox{fb$^{-1}$}}				%	fb^-1

\newcommand{\mum}{\mbox{$\mu$m}}				%	um

\newcommand{\tev}{\ensuremath{\mathrm{Te\kern -0.1em V}}}
\newcommand{\gev}{\ensuremath{\mathrm{Ge\kern -0.1em V}}}	%	GeV
\newcommand{\mev}{\ensuremath{\mathrm{Me\kern -0.1em V}}}	%	MeV
\newcommand{\kev}{\ensuremath{\mathrm{ke\kern -0.1em V}}}	%	keV
\newcommand{\massgev}{\mbox{\gev/$c^2$}}			%	GeV/c^2
\newcommand{\massmev}{\mbox{\mev/$c^2$}}			%	MeV/c^2
\newcommand{\pgev}{\mbox{\gev/$c$}}				%	GeV/c
\newcommand{\CP}{\ensuremath{\mathsf{CP}}}			%	CP

\newcommand{\pap}{\proton\antiproton}			%	ppbar
\newcommand{\proton}{\ensuremath{\mathrm{p}}}
\newcommand{\antiproton}{\ensuremath{\bar{\rm{p}}}}

\newcommand{\jpsi}{\ensuremath{J/\psi}}

\newcommand{\tab}[1]{tab.~\ref{tab:#1}}

\newcommand{\cita}[1]{\cite{#1}}

\newcommand{\dedx}{\ensuremath{\mathit{dE/dx}}}

\def\babar{\mbox{\slshape B\kern-0.1em{\smaller A}\kern-0.1em B\kern-0.1em{\smaller A\kern-0.2em R}}}
\begin{document}

\section{Spectroscopy at the Tevatron}

The theory of strong interactions (QCD) is a pristine example of the success of quantum field theory as a paradigm to describe Nature at the microscopic level. However, when the energy regime approaches $\mathcal{O}(100)$ MeV, the theory becomes non perturbative and calculations are challenging. This prevents QCD from predicting the spectrum of the physical hadron states and their properties. And, more generally,  it constrains seriously our understanding of the phenomenology of any process where soft QCD is involved, as transitions of heavy mesons sensitive to \CP-violation.  Numerical simulations in discretized space-time allow prediction of some quantities, and phenomenological \emph{effective} models provide additional information by exploiting decoupling of dynamical degrees of freedom. But accurate experimental input is crucial to  ensure adequate tuning of input parameters and constrain the approximations. The spectroscopy of heavy hadrons is particularly useful. It exploits the large mass difference between constituent quarks by approximating the heavy quark as a static color-field source for lighter partners, which  simplifies the calculations.  \par The CDF and D\O\ experiments at the Tevatron are major contributors,  owing to copious samples of all species of heavy hadrons available in $\sqrt{s}=1.96$ TeV \pap\ collisions, well-understood triggers and detectors, and advanced analysis techniques.  The experiments are similar, and focus on exclusive final states with charged particles. Since most recent result are from CDF, I only outline the CDF detector here. It is a multipurpose magnetic spectrometer surrounded by $4\pi$ calorimeters and muon detectors. Most relevant for heavy hadron spectroscopy are the tracking, particle-identification and muon detectors, and the trigger system.  
Six layers of double-readout silicon microstrip sensors between 2.5 and 22 cm from the beam,  and a single-readout layer at 1.5 cm radius,  provide precise vertex reconstruction, with approximately 15 (70)~\mum\  resolution in the azimuthal (longitudinal) direction.  A drift chamber  provides 96 samplings of three-dimensional charged-particles trajectories  between 40 and 140 cm radii in $|\eta|<1$,  for a transverse momentum 
resolution of   $\sigma_{p_{T}}/p_{T}^2 = 0.1\%/$(\pgev). Specific ionization measurements allow 1.5$\sigma$ separation between charged kaons (or protons) and pions, approximately  constant at momenta larger than 2~\pgev. A comparable identification is achieved at lower momenta by an array of scintillator bars at 140 cm radius, which  measure the time-of flight (TOF).   Muons with $p_T>1.5 (2.2)$ GeV/c are detected by planar  drift chambers at $|\eta|<0.6$ ($0.6<|\eta|<1.0$).  Low-$p_T$  dimuon triggers select  charmonium decays, totaling approximately 40 millions  $J/\psi$ mesons in 5~\lumifb\ of data.  These are used to reconstruct, for instance, 12~000 $X(3872)\to\jpsi\pi^+\pi^-$ decays.   A trigger on charged particles displaced from the primary vertex collects hadronic long-lived decays reconstructing tracks in the silicon with offline-like (48~\mum) impact parameter resolution, within 20 $\mu$s of the trigger latency. This yielded approximately 16~000 $\Lambda_b \to \Lambda_c^+\pi^-$  decays in 6~\lumifb\ of data. \par Both experiments collected 8~\lumifb\ of physics-quality data, which will reach 10~\lumifb\ by October 2011. Additional 6~\lumifb\ will be collected if the proposed three-year extension will be funded. \par Since the first Tevatron Run II publication \cita{1}, CDF and D\O\ have been steadily contributing to the understanding of heavy flavor spectroscopy. Both studied orbitally-excited mesons \cita{5}, observed exclusive $B_c^+$ decays \cita{3}, and studied exotic hadrons \cita{2}; CDF provided the best  measurements of $b$--hadron masses \cita{4} and has evidence of a new exotic state, the $Y(4140)$ \cita{Y4140}. Both experiments extended significantly the knowledge of $b$--baryons, with the first observations of $\Sigma_b^{\pm(*)}$ \cita{10}, $\Xi_b^{-}$ \cita{12}, and $\Omega_b^-$ \cita{23} families. I report recent, world-leading results on $\Sigma_b^{\pm(*)}$ and charmed baryons properties, and an updated analysis of $\jpsi\phi$ final states in search for exotic mesons. Throughout the text, the first uncertainty is statistical, the second one is systematic, and charge-conjugate states are implied.
\begin{figure}
\centering
\includegraphics[width=.45\textwidth]{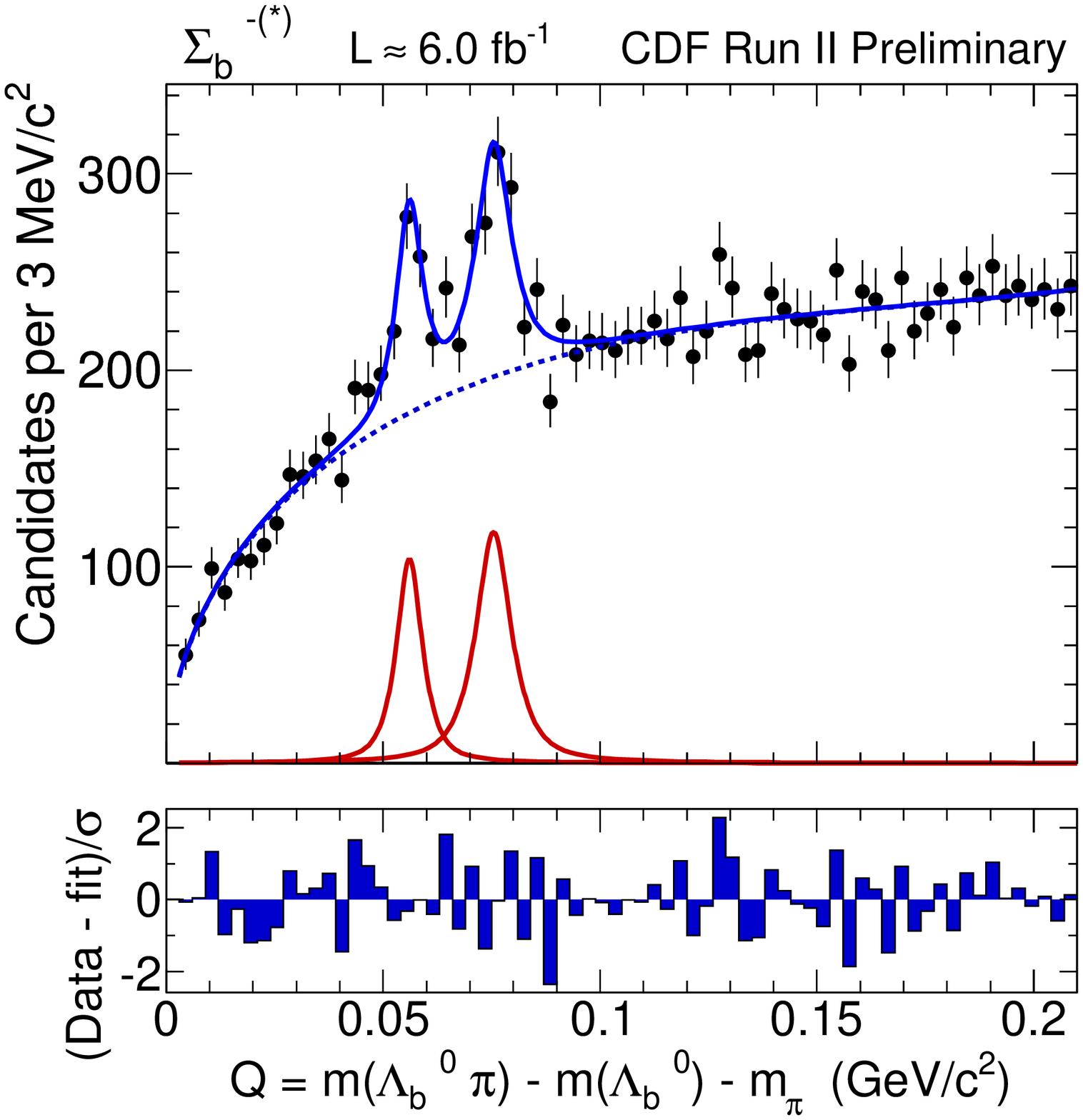}
\includegraphics[width=.45\textwidth]{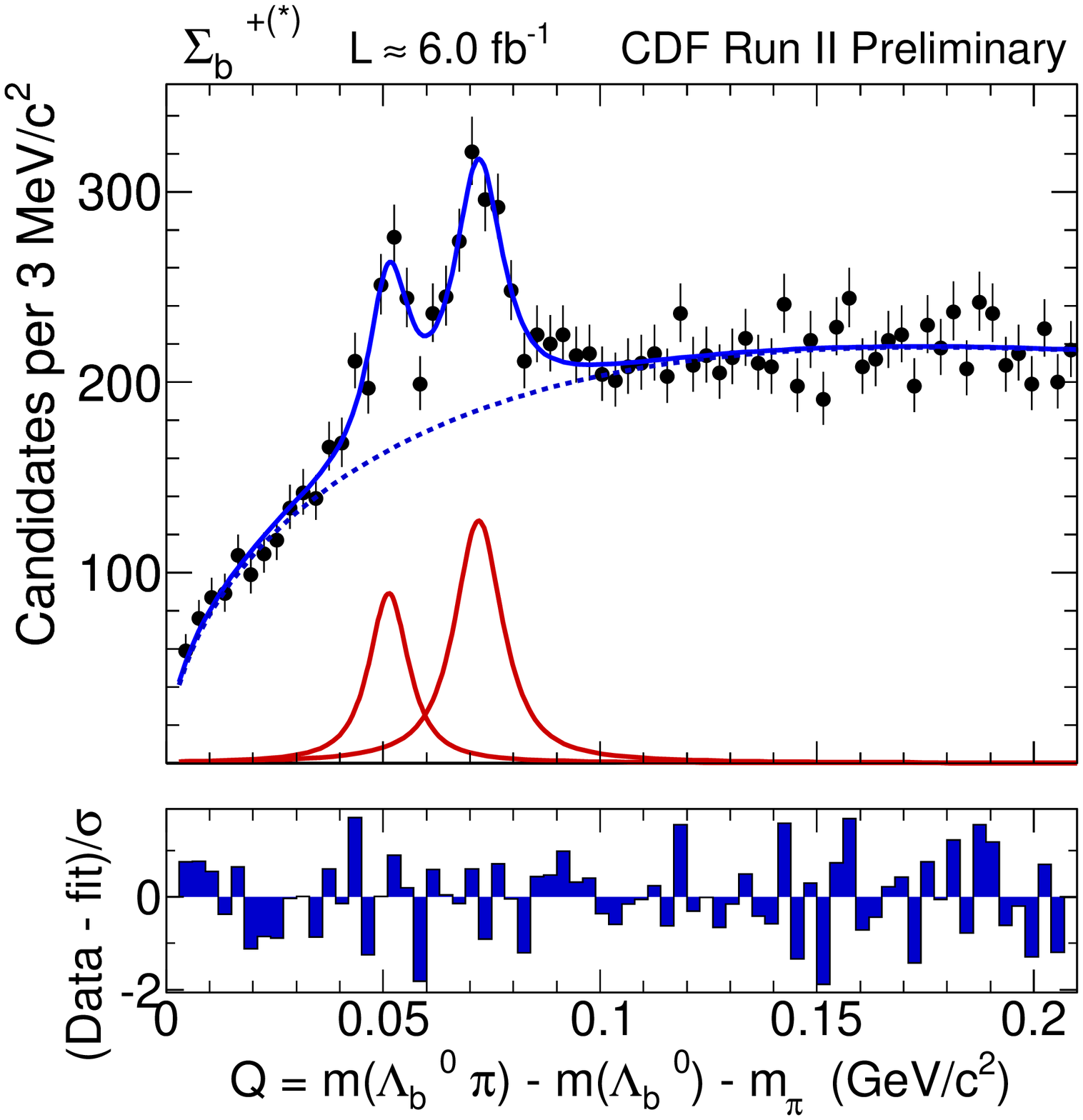}
\caption{$Q$-value distributions for $\Sigma_b^{-(*)}$ (left) and  $\Sigma_b^{+(*)}$ (right) candidates with fit overlaid.}
\label{fig1}
\end{figure}
\section{Improved measurement of $\Sigma_b^{\pm(*)}$ baryon resonance parameters}

The CDF experiment reports an update of an earlier measurement of  $\Sigma_b^{\pm(*)}$ (quark content $uub$ or $ddb$)  masses and widths using  $\Sigma_b^{\pm(*)} \to \Lambda_b(\to \Lambda_c^+\pi^-)\pi^{\pm}$ decays followed by $\Lambda_c\to \proton K^-\pi^+$, fully reconstructed in 6.0 fb$^{-1}$ of data collected by the hadronic trigger \cita{sigmapublic}. We reconstruct approximately 16~300 $\Lambda_b$ candidates with good purity  ($S/B\approx1.8$) using the known $\Lambda_c^+$ mass as a constraint in the four-tracks fit. The long $\Lambda_b$ lifetime is exploited by requiring at least two displaced tracks and the $\Lambda_b$ vertex being significantly displaced from the \pap\ vertex.
The $\Lambda_b$ candidates are then combined with a pion of $p_T > 0.2~\pgev$ into  $\Sigma_b^{\pm(*)}$ candidates.  
We study the $Q$-value distributions, $Q = m(\Lambda_b\pi^{\pm}) - m(\Lambda_b) - m(\pi^{\pm})$, to suppress the effects of $\Lambda_b$ mass  resolution (fig.\ref{fig1}). We fit independently  $\Sigma_b^{+(*)}$  and  $\Sigma_b^{-(*)}$ distributions using a non-relativistic Breit-Wigner function with $P$--wave-modified width, convoluted with two Gaussian distributions for detector resolution. The background is described using a second-degree polynomial modulated by a threshold $\sqrt{(Q+m_\pi)^2-\mathrm{thr}}$  function.  A Likelihood Ratio test (LR) against several null hypotheses yields statistical significances comfortably larger than $5.0\sigma$. The difference in yield between  $\Sigma_b^{+(*)}$ and  $\Sigma_b^{-(*)}$ signals (\tab{results}, top) is due to the charge-asymmetric tracking efficiency for low-momentum particles. The limited knowledge of tracking resolutions for soft tracks dominates the systematic uncertainties. We estimated a $20\%$ uncertainty on the widths by comparing the simulation with large samples of $D^{*+}\to D^0(\to K^-\pi^+)\pi^+$ decays, where the $D^{*+}$ and $\Sigma_b^{(*)\pm}$  low-momentum pions have similar kinematics. Smaller contributions are due to tracking momentum scale, slightly biased width estimates,  and the assumed models for signal and background. We determine masses with improved precision on the previous analysis (\tab{results}, top).  We also provide the first measurement of widths and isospin mass splittings within I = 1 triplets, with precision comparable with the one available for $\Sigma_c$ states: $m(\Sigma_b^+) - m(\Sigma_b^-)= -4.2_{-0.9-0.09}^{+1.1+0.07}~\massmev$, and $m(\Sigma_b^{*+}) - m(\Sigma_b^{*-}) = -3.0\pm0.9_{-0.13}^{+0.12}~\massmev$. 
\begin{table}[htb]
\begin{center}
\begin{tabular}{lccc}
\hline
\hline
State        & Mass [\massmev]  & Width [\massmev]      & Yield \\
%           & \massmev       & m, \massmev       & $\Gamma$, \massmev  & num. of cands. \\

\hline
\hline
%
%{$\Sigma_b^{+}$} & {\(52.0_{-0.8-0.4}^{+0.9+0.09} \)} & {\(5811.2_{-0.8}^{+0.9}\pm1.7 \)} & {\(9.2_{-2.9-1.1}^{+3.8+1.0} \)} & {\(468_{-95-15}^{+110+18} \)} \\
%{$\Sigma_b^-$} & {\(56.2_{-0.5-0.4}^{+0.6+0.07} \)} & {\(5815.5_{-0.5}^{+0.6}\pm1.7 \)} & {\(4.3_{-2.1-1.1}^{+3.1+1.0} \)} & {\(333_{-73}^{+93}\pm35 \)} \\
%{$\Sigma_b^{*+}$} & {\(72.7\pm0.7_{-0.6}^{+0.12} \)} & {\(5832.0\pm0.7\pm1.8 \)} & {\(10.4_{-2.2-1.2}^{+2.7+0.8} \)} & {\(782_{-103-27}^{+114+25} \)} \\
%{$\Sigma_b^{*-}$} & {\(75.7\pm0.6_{-0.6}^{+0.08} \)} & {\(5835.0\pm0.6\pm1.8 \)} & {\(6.4_{-1.8-1.1}^{+2.2+0.7} \)} & {\(522_{-76}^{+85}\pm29 \)} \\
%
{$\Sigma_b^{+}$} & {\(5811.2_{-0.8}^{+0.9}\pm1.7 \)} & {\(9.2_{-2.9-1.1}^{+3.8+1.0} \)} & {\(468_{-95-15}^{+110+18} \)} \\
{$\Sigma_b^-$} &  {\(5815.5_{-0.5}^{+0.6}\pm1.7 \)} & {\(4.3_{-2.1-1.1}^{+3.1+1.0} \)} & {\(333_{-73}^{+93}\pm35 \)} \\
{$\Sigma_b^{*+}$} & {\(5832.0\pm0.7\pm1.8 \)} & {\(10.4_{-2.2-1.2}^{+2.7+0.8} \)} & {\(782_{-103-27}^{+114+25} \)} \\
{$\Sigma_b^{*-}$} & {\(5835.0\pm0.6\pm1.8 \)} & {\(6.4_{-1.8-1.1}^{+2.2+0.7} \)} & {\(522_{-76}^{+85}\pm29 \)} \\
\hline
{$\Sigma_c(2455)^{++}$} & {\(2453.90 \pm 0.13 \pm 0.14 \)} & {\(2.34 \pm 0.13 \pm 0.45 \)} & {\(\approx 13~800\)} \\
{$\Sigma_c(2455)^0$} & {\(2453.74 \pm 0.12 \pm 0.14 \)} & {\(1.65 \pm 0.11 \pm 0.49 \)} & {\(\approx 15~900\)} \\
{$\Sigma_c(2520)^{++}$} & {\(2517.19 \pm 0.46 \pm 0.14 \)} & {\(15.03 \pm 2.12 \pm 1.36 \)} & {\(\approx 8~800 \)} \\
{$\Sigma_c(2520)^{0}$} & {\(2519.34 \pm 0.58 \pm 0.14 \)} & {\(12.51 \pm 1.82 \pm 1.37 \)} & {\(\approx 9~000 \)} \\
{$\Lambda_c(2595)^{+}$} &{\(2592.25 \pm 0.24 \pm 0.14 \)} & {\(2.59 \pm 0.30 \pm 0.47\)} & {\(\approx 3~500\)} \\
{$\Lambda_c(2625)^{+}$} &{\(2628.11 \pm 0.13 \pm 0.14 \)}  & {\(<0.97  \mbox{ at the 90\% CL} \)} & {\(\approx 6~200 \)} \\
\hline
\hline
%
%  & \multicolumn{3}{c}{Isospin Mass Splitting, \massmev}  \\
%\hline
%
 %{m($\Sigma_b^+$) - m($\Sigma_b^-$)}  & \multicolumn{4}{c}{\( -4.2_{-0.9-0.09}^{+1.1+0.07} \)}\\
 %{m($\Sigma_b^{*+}$) - m($\Sigma_b^{*-}$)} & \multicolumn{4}{c}{\(-3.0\pm0.9_{-0.13}^{+0.12} \) }\\
%
%\hline
%\hline
%
\end{tabular}
\caption{Summary of baryon results.}
\label{tab:results}
\end{center}
\end{table}
\section{Study of $\Lambda_c(2595)^+$, $\Lambda_c(2625)^+$, $\Sigma_c(2455)^{++,0}$ and $\Sigma_c(2520)^{++,0}$  charmed baryons}
\begin{figure}
\centering
\includegraphics[width=.45\textwidth]{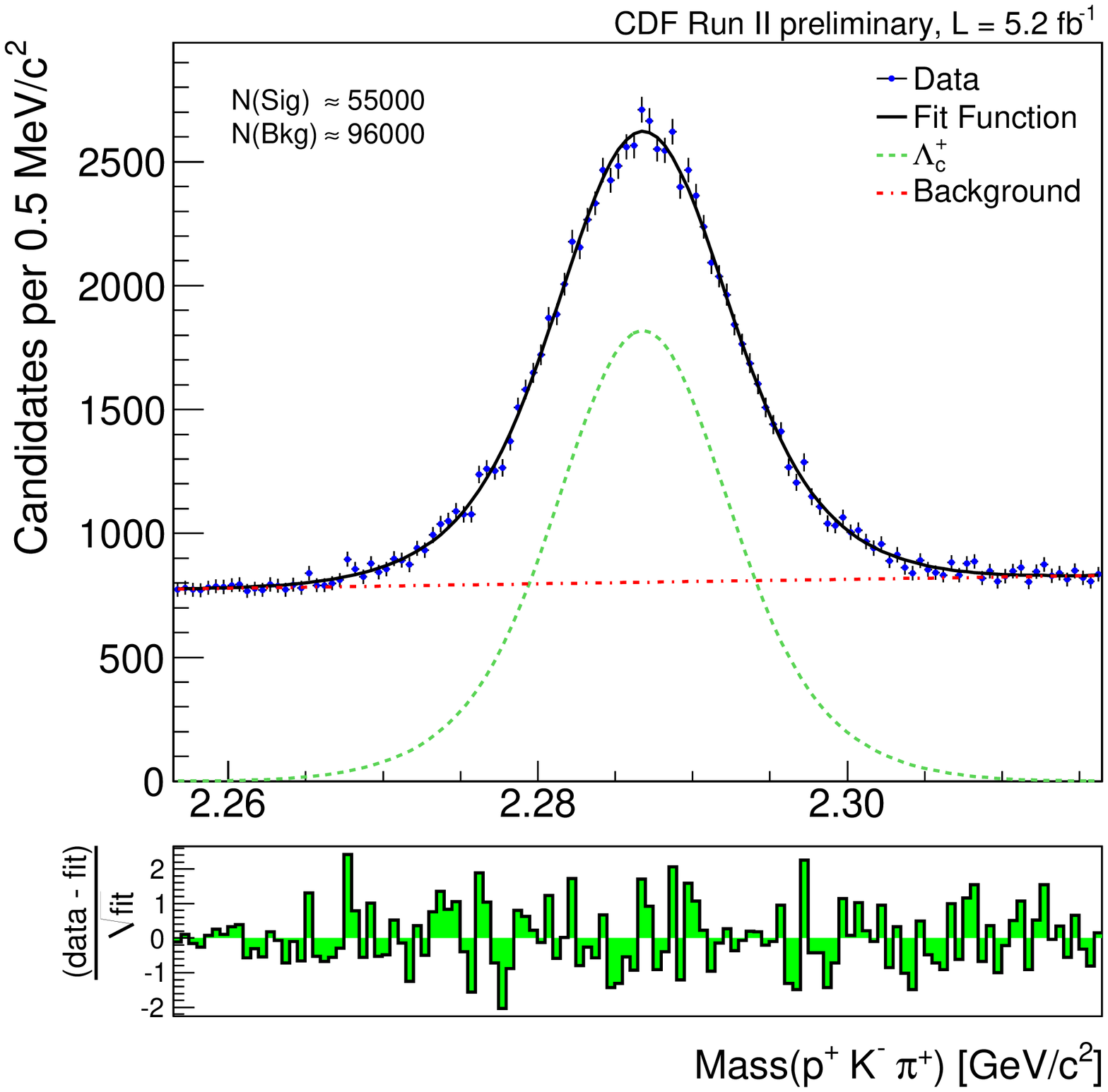}
\includegraphics[width=.45\textwidth]{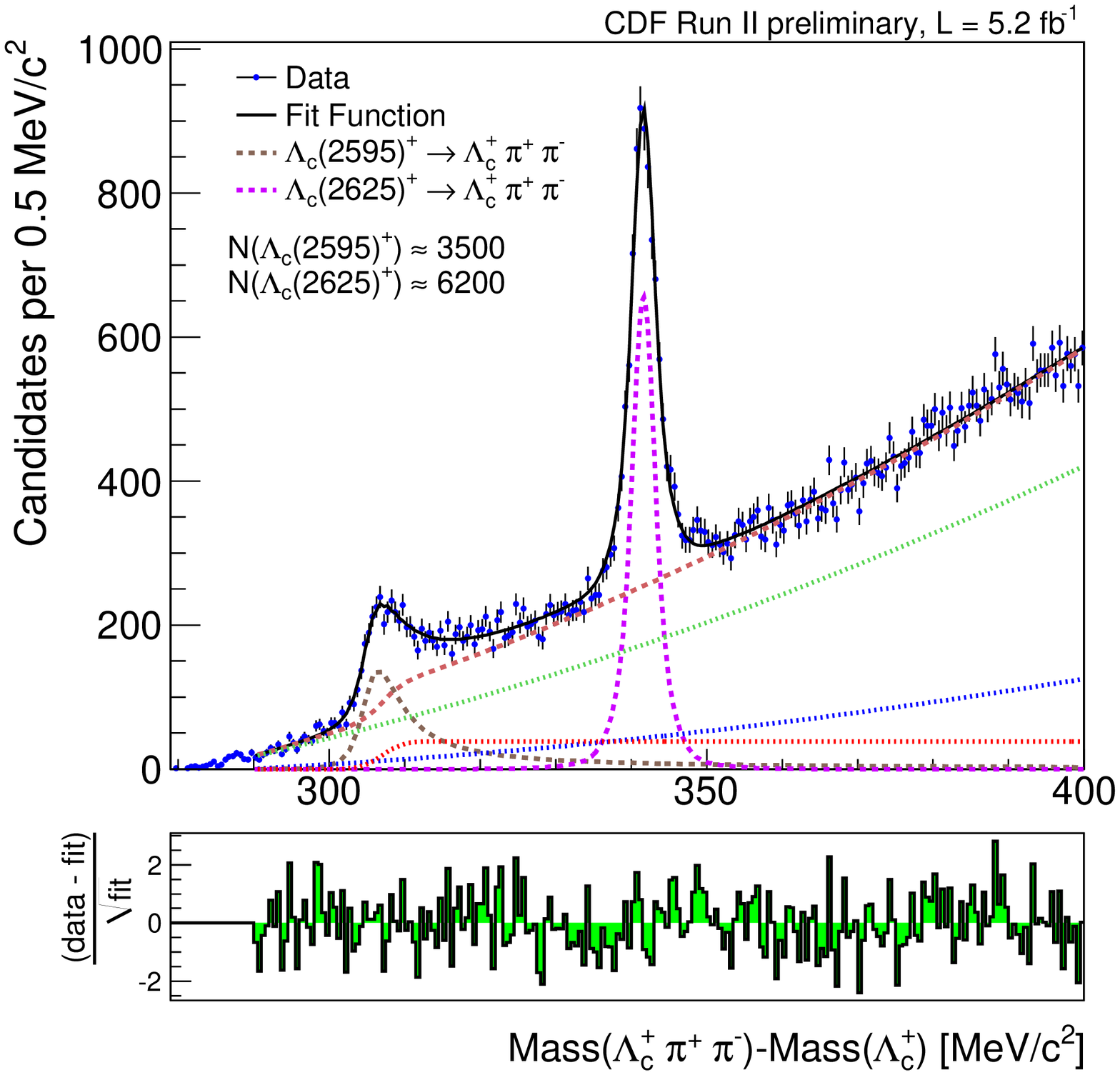}
\includegraphics[width=.45\textwidth]{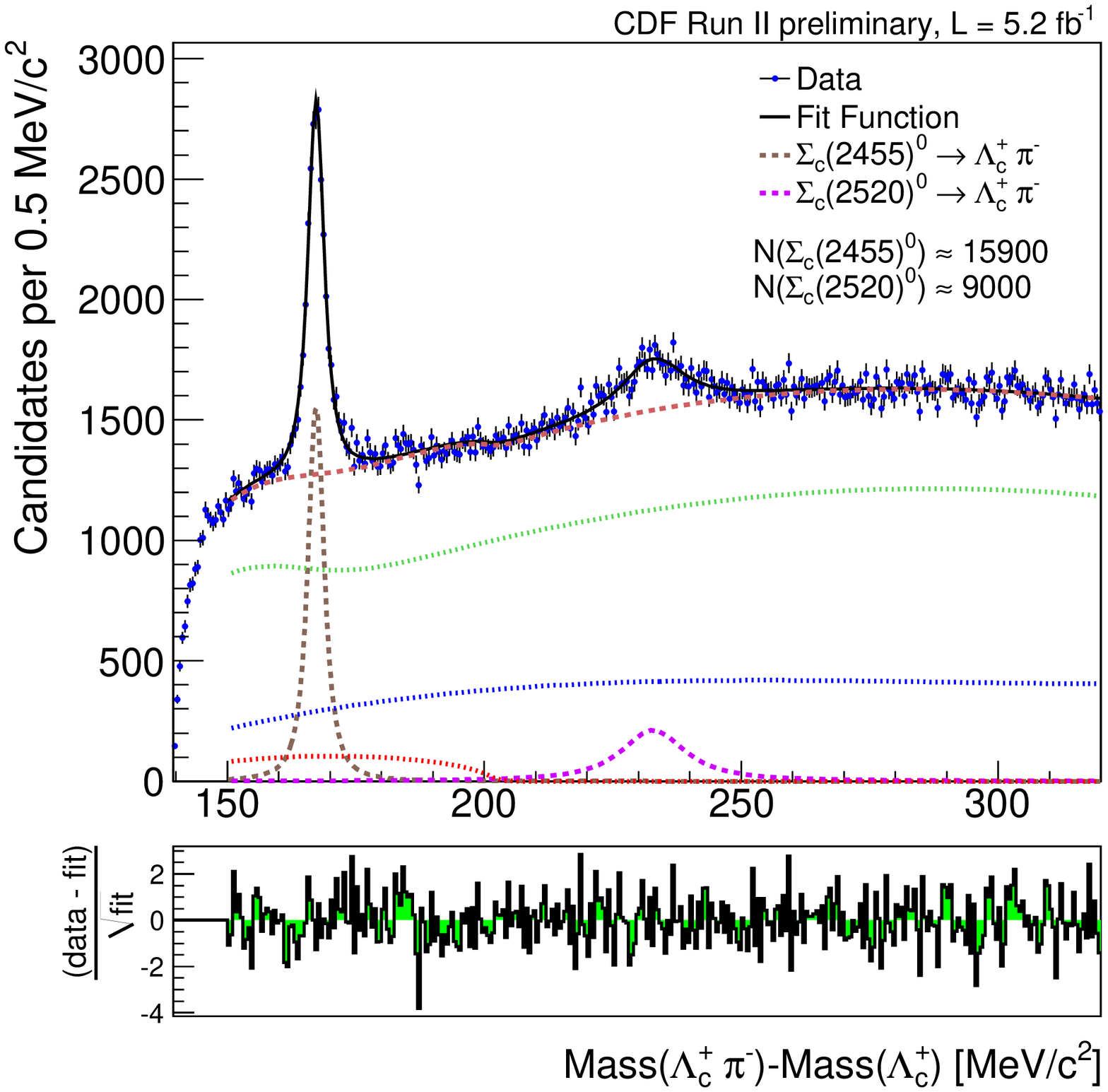}
\includegraphics[width=.45\textwidth]{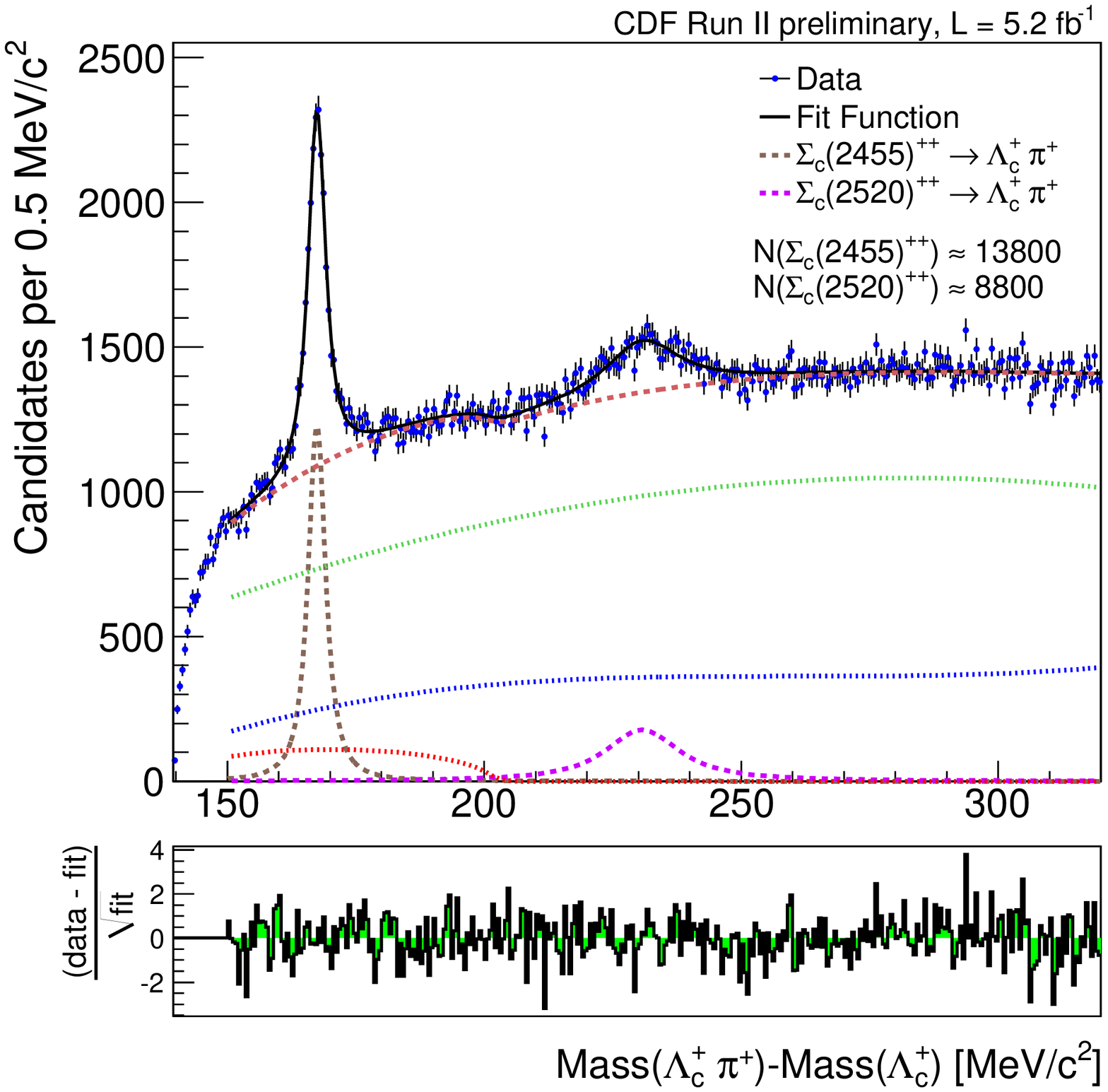}
\caption{Mass distribution of $\Lambda_c^+$ candidates (top left). Mass difference for  $\Lambda_c^{+*}$ (top right) $\Sigma_c^0$ (bottom left),  $\Sigma_c^{++}$ (bottom right). Fit results are overlaid.}
\label{fig2}
\end{figure}

In a new analysis of charmed baryons, CDF measure masses and widths of $\Lambda_c(2595)^+$, $\Lambda_c(2625)^+$, $\Sigma_c(2455)^{++,0}$ and $\Sigma_c(2520)^{++,0}$ states \cita{charmpublic}.
We study the largest available samples of $\Sigma_c$ states in $\Lambda_c\pi$ decays, and $\Lambda_c^{+*}$ in $\Lambda_c^+\pi^+\pi^-$ final states, through intermediate $\Sigma_c$ resonances. Cross-contamination among channels requires careful treatment of the background because different sources populate different phase space regions. We exploit a sample of $\Lambda_c^+\to \proton K^-\pi^+$ decays reconstructed in  5.2 fb$^{-1}$ collected by the displaced track trigger. An artificial neural network classifier (NN) selects a pure subsample of approximately 55~000 $\Lambda_c^+$ decays (fig.~\ref{fig2}, top left) and another one optimizes the reconstruction of each signal from the $\Lambda_c^+$ candidates. Both NN are trained using data only. Masses and widths are measured through binned likelihood fits of distributions of masses subtracted by the $\Lambda_c^+$ mass, $\Delta M$,
to suppress effects of tracking resolution (fig.\ref{fig2}).   The $\Sigma_c(2455)^{++,0}$ and  $\Sigma_c(2520)^{++,0}$ states are studied through the mass differences of $\Lambda_c^+ \pi^+$ and $\Lambda_c^+ \pi^-$. The distribution of $\Lambda_c^+\pi^+\pi^-$ is used  for the $\Lambda_c(2595)^+$ and $\Lambda_c(2625)^+$.  Non-relativistic Breit-Wigner functions model the $\Sigma_c$ states. For the $\Lambda_c(2595)^+$ we use a mass-dependent width to correct for the proximity of the $\Sigma_c(2455)\pi$ threshold and remove biases of the mass toward higher values \cite{blechman}. The Breit-Wigner are convoluted with resolution functions from simulation, expanded in sum of Gaussians centered in zero with 1.6--2.6~\massmev\ mean widths, depending on the channel. The background is dominated by random combinations of tracks without a real $\Lambda_c^+$, which are modeled with empiric fits to the $\Delta M$ distribution of events from $\Lambda_c^+$ mass sidebands. Real $\Lambda_c^+$ combined with random tracks are modeled with polynomial functions determined by the fit. Non-resonant $\Lambda_c(2625)^+$ decays (in the $\Sigma_c$ case) and real $\Sigma_c$ combined with random tracks  ($\Lambda_c^{+*}$) also contribute.
%The difference between doubly-charged and neutral $\Sigma_c$ spectra combinations is due to $D^(2010)+ mesons with multibody D0 decays, where not all D0 decay products are reconstructed. In order to describe this reflection, an additional Gaussian function is used. 
Detector resolution at low-momentum drives the systematic uncertainties. As in the $\Sigma_b^{\pm(*)}$ case, this is constrained by using  $D^{*+}$ and $\psi(2S)\to J/\psi(\to \mu^+\mu^-)\pi^+\pi^-$ decays. Finite knowledge of the momentum scale, assumptions in the fit model and uncertainties on the external inputs to the fit add further contributions. All results (\tab{results}, bottom) agree with previous determinations except for the $\Lambda_c(2595)^+$ mass. Detailed treatment of threshold effects shifts it by 3.1~\massmev\  from the currently known value. We derive  a 90\% credibility Bayesian upper limit assuming a uniform positively-defined prior for the $\Lambda_c(2625)^+$ width because it is consistent with zero. The value $h_2^2=0.36\pm0.04\pm0.07$ for the squared pion decay-constant  is derived from the $\Lambda_c(2595)^+$ width \cite{blechman}. The precision of all results is comparable or superior to that of existing measurements.

\section{Updated search for exotic $\jpsi\phi$ states}

\begin{figure}
\centering
\includegraphics[width=.45\textwidth]{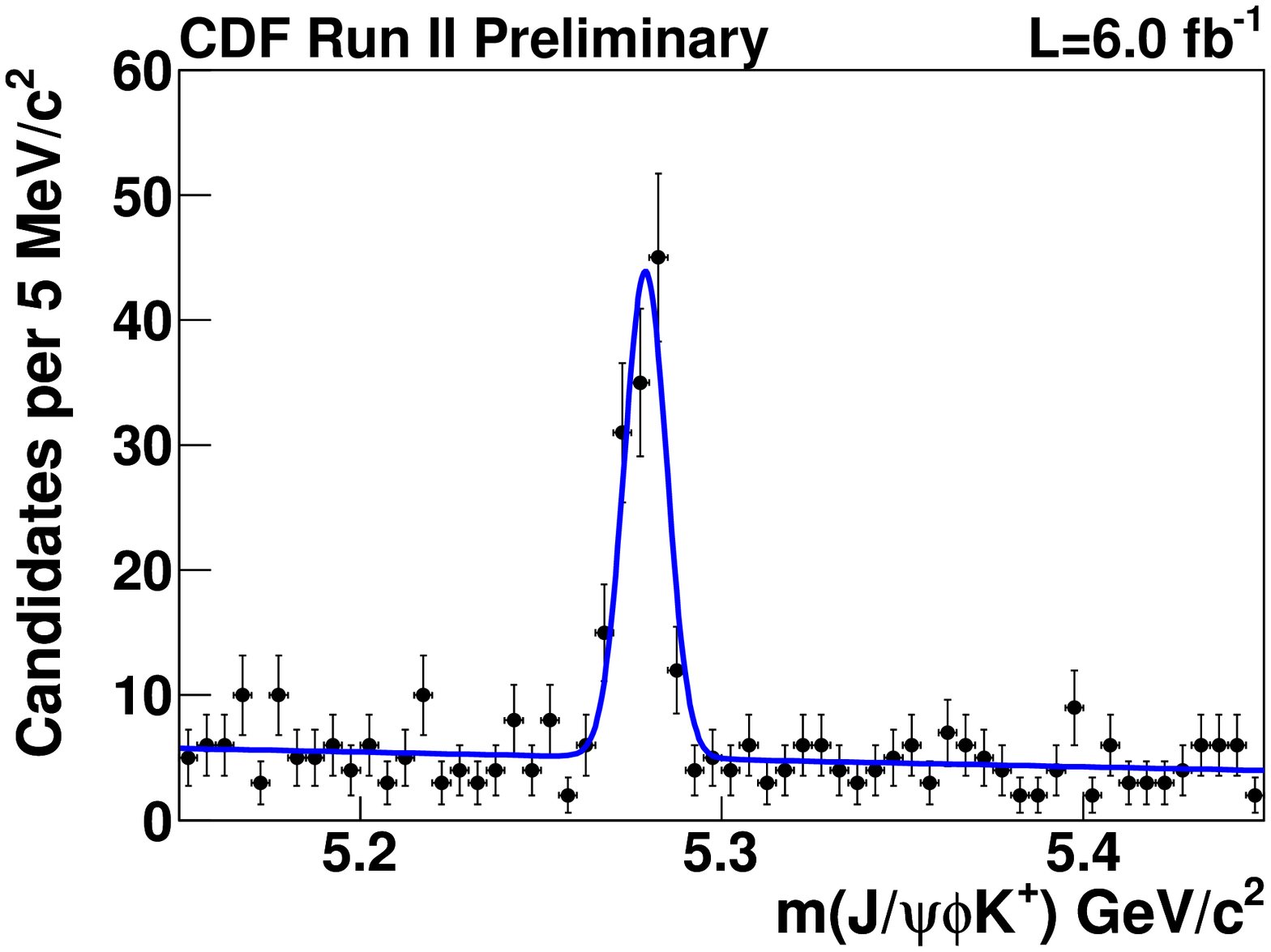}
\includegraphics[width=.45\textwidth]{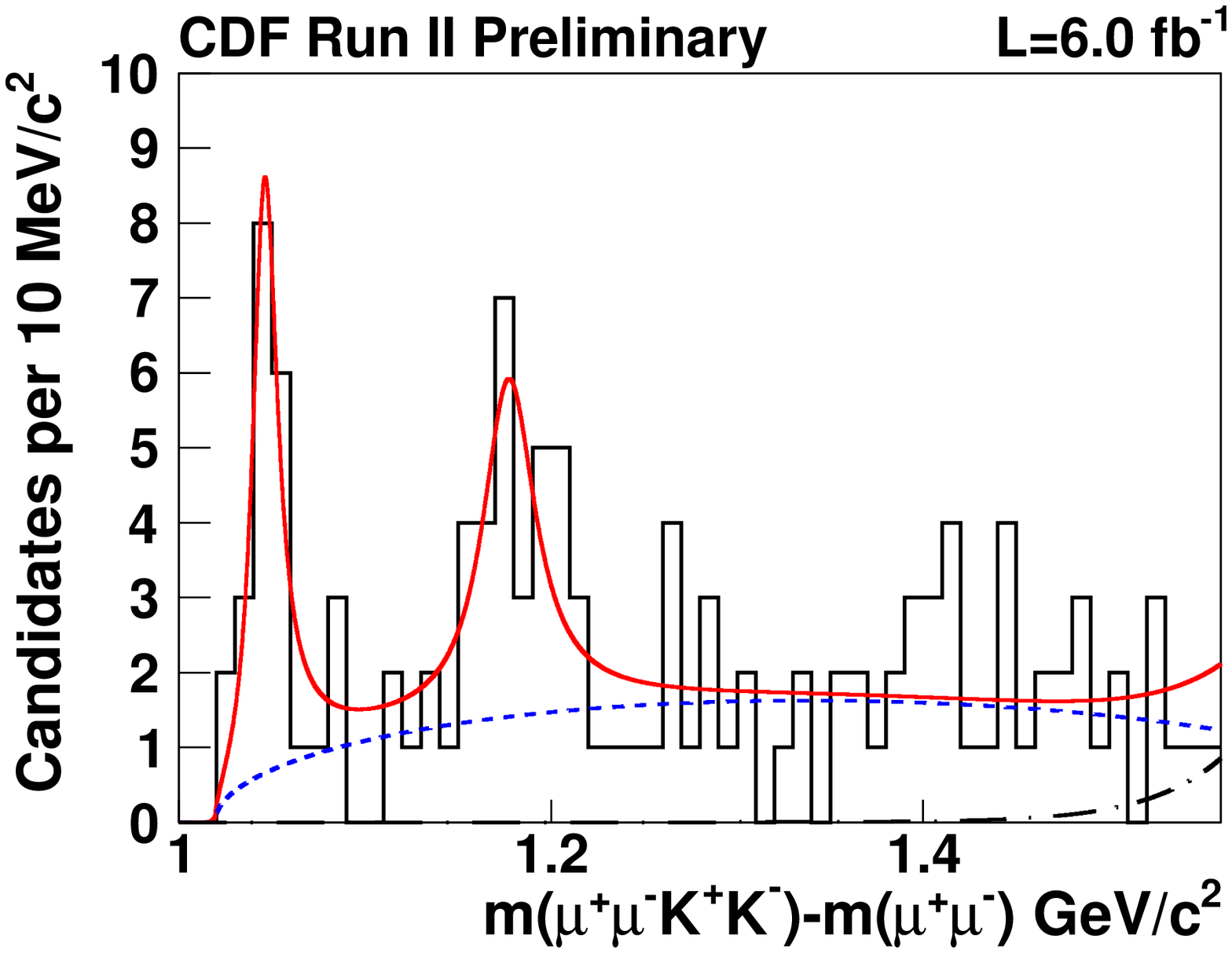}
\caption{Distribution of $\jpsi\phi K^+$ mass (left) and $\Delta m$ for events in the $B^+$ signal (right) with fit overlaid.}
\label{fig3}
\end{figure}

In 2009, CDF  had an evidence for a $\jpsi\phi$ structure,  dubbed $Y(4140)$,  in exclusive $B^+\to \jpsi\phi K^+$ decays reconstructed in 2.7~\lumifb\ collected by the dimuon trigger \cita{Y4140}. %CDF contributed to the plethora of XYZ states with charmonium-like decays that may not   fit the charmonium picture discovered in the recent years. Interpretations for the nature of these states include quark-gluon-hybrids, diquark-diantiquark bound states, molecular states composed of pairs of usual mesons, glueballs, and so forth. 
Belle and Babar, with lower acceptance for low-momentum charged particles, could not confirm or disprove. We report an update on 6.0~\lumifb\ with no changes in the candidate selection \cita{ypublic}.  Candidates compatible with $\jpsi\to\mu^+\mu^-$ and $\phi \to K^+K^-$ decays are combined with a charged particle with kaon mass assignment in a kinematic fit to a displaced vertex.  Vertex displacement and kaon identification (\dedx\ and TOF) reduce the background by about five orders of magnitude. The combinatorial component is further reduced by requirements of the narrow $B^+$ mass window. Figure \ref{fig3}, left,  shows the mass spectrum where a signal of $115\pm12$ $B^+$ decays is visible. Despite the two-fold increase in integrated luminosity,  this is only a 53\% increase in signal yield over the previous analysis, because of trigger bandwidth limitation.  %Retaining candidates with mass within $\pm3\sigma$ ($\sigma \approx 18~\massmev$) around the known $B^+$ mass are retained. %Furthermore, sideband events within [?9, ?6]? and [+6, +9]? around the nominal B+ mass are used to reflect the combinatorial background in the J/?? spectrum.
A fit of the $K^+K^-$ spectrum show no appreciable contribution from $f_0(980)$ or non-resonant $K^+K^-$ decays. The $Y(4140)$ appears as a narrow near-threshold excess in the distribution of mass difference  $\Delta m = m(\mu^+\mu^- K^+ K^-)-m(\mu^+\mu^-)$ (fig~\ref{fig3}, right). No structure is visible in the same spectrum sampled in $B^+$ mass sidebands. We exclude events with $\Delta m > 1.56~\massgev$ to reject  background from misidentified $B^0_s \to  \psi(2S) \phi \to (\jpsi \pi^+ \pi^-)\phi$ decays, which are the only potential contaminants, as resulting from a simulation of all $B$ decays with a \jpsi\ in the final state. We describe the enhancement as the convolution of an $S$-wave relativistic Breit-Wigner with a Gaussian resolution of $1.7~\massmev$ (from simulation) in an unbinned likelihood fit. We assume the background distributed as a three-body phase-space. We obtain a signal of $19\pm6\ \pm 3$ events with mass $m = 4143.4 ^{+2.9}_{-3.0} \pm0.6~\massmev$ and width $\Gamma= 15.3^{+10.4}_{-6.1}\pm2.5~\massmev$, consistent with the previous analysis. The size of the width suggests a strong-interaction decay. The branching fraction relative to the non-resonant $B^+\to \jpsi\ \phi K^+$ decay is $0.149\pm0.039\pm 0.024$, with assumptions on the $Y(4140)$ $J^{PC}$ quantum numbers. The significance is determined from the LR distribution in statistical  trials of three-body phase space $\Delta m$ distributions generated to mimic background. Only two out of ten millions trials show a structure with LR equal or higher than observed in data, corresponding to a statistical significance of 5.0$\sigma$. An additional excess appears at $\Delta m\approx 1.18~\massmev$.  Assuming the existence of the $Y(4140)$ and a three-body phase-space background, this enhancement is fit using a  $3.0~\massmev$ Gaussian detector resolution. This yields a signal of $22\pm8$ signal events, located at $m = 4274.4^{+8.4}_{-6.7}~\massmev$ mass, with $\Gamma = 32.3^{+21.9}_{-15.3}~\massmev$ width and a significance of  $3.1\sigma$.

\end{document}